\documentclass{emulateapj}  

\usepackage{graphicx} 
\usepackage{apjfonts} 
\usepackage{epsfig,color,colordvi}

\newcommand{\ia}{\'{\i}}

\shorttitle{}
\shortauthors{Ortiz et al.}

\begin{document}

\title{Emergence of granular-sized magnetic bubbles through the solar atmosphere. III. The path to the transition region.}

\author{Ada Ortiz$^{1,2}$}
\affil{$^1$Institute of Theoretical Astrophysics, University of Oslo, P.O. Box 1029 Blindern, N-0315 Oslo, Norway}
\affil{$^2$Instituto de Astrof\ia sica de Andaluc\ia a (CSIC), Apdo. 3040, 18080 Granada, Spain}
\email{ada@astro.uio.no} 
\author{Viggo H. Hansteen}
\affil{Institute of Theoretical Astrophysics, University of Oslo, P.O. Box 1029 Blindern, N-0315 Oslo, Norway} 
\author{Luis Ram\'on Bellot Rubio}
\affil{Instituto de Astrof\ia sica de Andaluc\ia a (CSIC), Apdo. 3040, 18080 Granada, Spain} 
\author{Jaime de la Cruz Rodr\ia guez}
\affil{Institute for Solar Physics, Dept. of Astronomy, Stockholm University, Albanova University Center, SE-10691 Stockholm, Sweden}
\author{Bart De Pontieu$^{3,1}$}
\affil{$^3$Lockheed Martin Solar and Astrophysics Laboratory, 3251 Hanover Street, Org. A021S, Bldg. 252, Palo Alto, CA 94304, USA} 
\affil{$^1$Institute of Theoretical Astrophysics, University of Oslo, P.O. Box 1029 Blindern, N-0315 Oslo, Norway}
\author{Mats Carlsson}
\author{Luc Rouppe van der Voort} 
\affil{Institute of Theoretical Astrophysics, University of Oslo, P.O. Box 1029 Blindern, N-0315 Oslo, Norway}

\begin{abstract}

We study the ascent of granular-sized magnetic bubbles from the solar photosphere through the chromosphere into the transition region and above, for the first time. Such events occurred in a flux emerging region in NOAA 11850 on September 25, 2013. During that time, the first co-observing campaign between the Swedish 1-m Solar Telescope and the IRIS spacecraft was carried out. Simultaneous observations of the chromospheric H$\alpha$ 656.28 nm and \ion{Ca}{2} 854.2 nm lines, plus the photospheric \ion{Fe}{1} 630.25 nm line, were made with the CRISP spectropolarimeter at the SST reaching a spatial resolution of 0."14. At the same time, IRIS was performing a four-step dense raster of the said emerging flux region, taking slit-jaw images at 133 (C~{\sc ii}, transition region),
140 (\ion{Si}{4}, transition region), 279.6 (\ion{Mg}{2} k, core, upper chromosphere), and 283.2 nm (\ion{Mg}{2} k, wing, photosphere). Spectroscopy of several lines was performed by the IRIS spectrograph in the far and near ultraviolet, of which we have used the \ion{Si}{4} 140.3 and the \ion{Mg}{2} k 279.6 nm lines. Coronal images from the Atmospheric Imaging Assembly of the Solar Dynamics Observatory were used to investigate the possible coronal signatures of the flux emergence events. The photospheric and chromospheric properties of small-scale emerging magnetic bubbles have been described in
detail in Ortiz et al. (2014). Here we are able to follow such structures up to the transition region. We describe the properties, including temporal delays, of the observed flux emergence in all layers. We believe this may be an important mechanism of transporting energy and magnetic flux from subsurface layers to the transition region and corona. \\

\end{abstract}


\section{Introduction}

An important facet of solar activity is the emergence of magnetic flux through the photosphere and into the
regions above. This process occurs on many scales, from the size of a granule to full active regions spanning several $100\times
100$~arcsec$^2$ (see reviews by \citet{2014SSRv..186..227S} and \citet[]{2014LRSP...11....3C}). 

Initially magnetic flux rises from deep within the convection zone before stalling just below the photosphere, where it is
no longer buoyant. If and when the field manages to burst through the photosphere, it rapidly expands into the outer atmosphere filling a considerable volume horizontally \citep{2004A&A...426.1047A}. 
Observations indicate that the emerging flux produces a cold bubble that interacts with the
ambient field and tends to push the transition region and corona aside, thus extending the relatively cool chromospheric plasma several Mm above the photosphere \citep{2014ApJ...781..126O,2015ApJ...810..145D}. Simulations by \citet{2008ApJ...679..871M,2009ApJ...702..129M,2009A&A...507..949T,2014ApJ...788L...2A} portray a similar picture.


A miscellany of dynamical phenomena such as surges, CMEs, flares, Ellerman bombs and jets, to mention a few, occur when the newly emerged field interacts with the pre-existing (so-called ambient) field \citep{2012ASPC..455..109G,2014SSRv..186..227S,2014LRSP...11....3C}. Magnetic reconnection between these two flux systems is thought to be the main cause behind such an energy release and dynamical phenomena \citep{1977ApJ...216..123H}. 

According to \citet{2004ApJ...614.1099P}, 
flux emerges into the outer atmosphere through the photosphere in the form of a serpentine field due to the undular instability. Plasma flows in the troughs of the undular field lines forming `U-shaped' loops, which can produce reconnection already at the photospheric level (hypothetically producing Ellerman bombs) as the field is squeezed by granular, convective motions. \citet{2004ApJ...614.1099P} argue that the field relieved of its mass load by reconnection is free to expand upwards, and there to form longer loops, perhaps being the source of H$\alpha$ fibrils and/or transition region and coronal loops with lengths of the same order as the emerging active region. Anchored to the active region, these longer loops constitute the environment into which newly formed magnetic bubbles expand.

The interaction between the new and the pre-existing field depends both on the {\it strengths} of the fields involved and on the relative {\it angle} between them \citep{2004A&A...426.1047A,2007ApJ...666..516G}. 
A large variety of dynamic phenomena and transient heating may occur, generally with more violent events associated with larger angles of interaction. 

Another important factor is {\it height}: the interaction of new emerging loops with ambient fields at different altitudes produces different transient phenomena \citep{2007ApJ...657L..53I}. 
For example, Ellerman bombs are produced at low altitudes while X-ray jets happen when the interaction occurs at greater heights. Surges would be ejections of chromospheric plasma with a filamentary structure that may occur due to plasma squeezing in the horizontal expansion of the loops \citep{1993ASPC...46..507K}. 
Many observations show that surges are often correlated with EUV coronal jets \citep{2008A&A...481L..57C} 
and with X-rays jets with a maximum velocity of 200 km/s \citep{1992PASJ...44L.173S}. In summary, the relationship between flux emergence and such dynamic phenomena as Ellerman bombs, surges, and jets is well established \citep[see e.g. a review by][and references cited therein]{2012ASPC..455..109G}. 

Very detailed observations of small-scale flux emergence at high resolution have hitherto been restricted to the photosphere and the chromosphere \citep[e.g.][henceforth Paper~\sc{i}]{2010ApJ...724.1083G,2014ApJ...781..126O}. In spite of recent developments, however,  there is still a lack of connectivity between all the layers of the solar atmosphere {\it at very high resolution}. Most flux emergence studies either concentrate on one layer of the atmosphere or do not have the appropriate spatial, spectral or temporal resolution. In order to scan all layers of the Sun and follow the magnetic field on its way up, one needs {\it simultaneous multi-wavelength} observations at the highest possible resolution. Currently, this is only feasible by combining observations from different telescopes and coping with the inherent difficulties of multi-instrument data sets.

\citet{2010ApJ...724.1083G} is an example of a detailed multi-wavelength analysis of a small-scale flux emergence event. Using data from the Hinode spacecraft and the Swedish 1-meter Solar Telescope they followed the emergence of flux from the photosphere to the corona. Unfortunately, their observations had a moderate spatial resolution of 1\arcsec  in the transition region with the Hinode/EIS instrument \citep{2007SoPh..243...19C} and less resolution with the coronal XRT imager \citep{2007SoPh..243...63G}. \citet{2010ApJ...724.1083G} found that intensity enhancements occur when the emerging footpoints meet pre-existing flux of opposite polarity. Such enhancements are first seen in the lower chromosphere, then in the upper chromosphere and finally in the corona. 

In this third paper of our flux emergence series we combine observations at the highest spatial, spectral and temporal resolution of small-scale magnetic bubbles rising from the photosphere to the transition region, both with imaging and with spectroscopic information in each layer of the atmosphere. We believe this is the first time a single flux emergence event is followed up to the transition region with spectroscopy at a spatial resolution of at least 0."33 and enough cadence to capture the fast dynamics of its interactions with the chromosphere and transition region. In order to achieve these goals we count on simultaneous multi-wavelength observations obtained with the SST, the IRIS spacecraft and the SDO spacecraft.


In Paper I of this series we described small-scale flux emergence in the photosphere, and found that it could be best described as the rise of a cold, magnetized bubble, seen as a dark feature in the wings of the \ion{Ca}{2}~854.2 nm line. This was an observational confirmation of the bubbles that had been predicted by \citet{2008ApJ...679..871M,2009ApJ...702..129M}  and \citet{2009A&A...507..949T} from simulations of flux emergence. A detailed characterization of these bubbles, concentrating on their temporal evolution, morphological, dynamic and magnetic properties from the photosphere and up to the mid-chromosphere was carried out in Paper I. In addition, a comparison with a realistic 3D MHD numerical simulation that shows similar features was performed. In \citet[][hereafter called Paper~{\sc ii}]{2015ApJ...810..145D} this study was continued with particular attention paid to understanding the peculiar shape of the \ion{Ca}{2}~854.2~nm intensity profiles observed within the flux emergence region. Using non-LTE inversions and comparisons with numerical simulations, the structuring of the physical conditions of the photosphere and low-to-mid chromosphere were investigated during the emergence of a magnetic bubble.


In this paper we follow the evolution of two granular-sized emergence events in a newly forming active region as they rise through the photosphere, chromosphere and interact with the transition region. Imaging and spectroscopic information will give us the signatures of the rising plasma at several heights: lifetimes of the magnetic bubbles, temporal delays between the passage through different points and vertical velocities. 

In section~\ref{data} we describe the observations and the data analysis. In section~\ref{results} we present the results obtained for the two flux emergence events. Finally, section~\ref{disc} summarizes our findings and possible interpretations.

\section{Observations and data analysis}
\label{data}


The first coordinated campaign between the Interface Region Imaging Spectrograph \citep[IRIS;][]{2014SoPh..289.2733D} and the Swedish 1-meter Solar Telescope \citep[SST;][]{2003SPIE.4853..341S} was carried out in August and September 2013. Combining simultaneous SST and IRIS observations of flux emergence allows one to follow this fundamental process through the entire atmosphere up towards coronal temperatures. 

\subsection{Observations}
\label{obs}

The IRIS dataset was acquired on September 25, 2013 starting at  08:09:43 UT. This is a four-step dense raster
with steps of $0\farcs35$. The exposure time was 2~s for each slit position and the raster cadence 11.6~s. The observations cover a field-of-view of $1 \arcsec \times 60 \arcsec$ centered at $(x,y)=(-58.8, 37.9) \arcsec$ or $\theta=4^{\circ}$, in NOAA AR 11850. The slit was rotated 90 degrees with respect to solar north and solar rotation compensation was not switched on. The time series comprises 600 rasters and ended at 10:05:32 UT.
At the same time, slit-jaw images were taken in four spectral bands: at 
$133\,$nm (FWHM 5.5~nm, dominated by \ion{C}{2} lines and continuum), 
$140\,$nm (FWHM 5.5~nm, dominated by \ion{Si}{4} lines and continuum), 
and $279.6\,$nm (FWHM 0.4~nm, centered on \ion{Mg}{2}~k) with 11.6~s cadence 
and at $283.2\,$nm (FWHM 0.4~nm, positioned in the far \ion{Mg}{2}~h wing) with 69.5~s cadence.

\begin{figure}
\hspace{-1.5cm}
\resizebox{1.25\hsize}{!}{\includegraphics{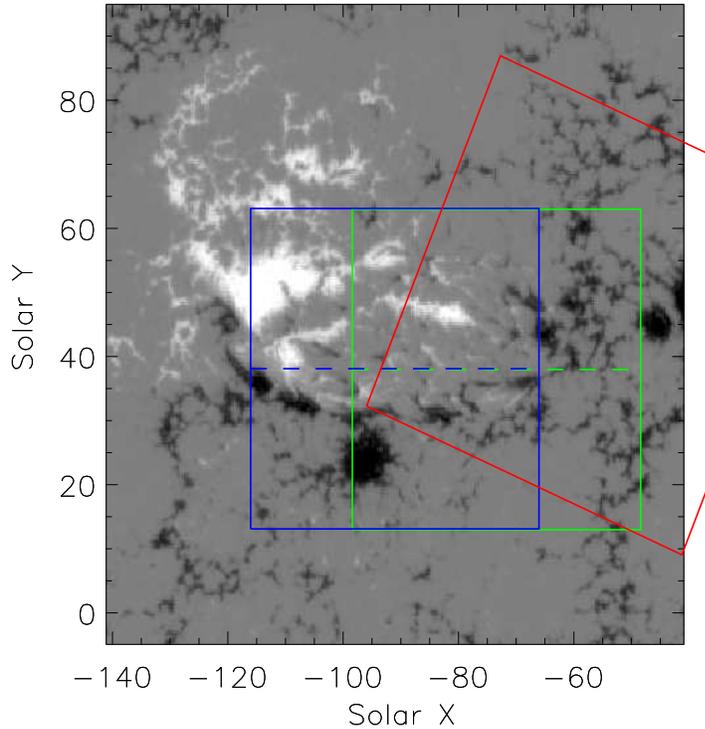}}
\caption{Context Hinode/SP magnetogram taken on September 25, 2013 at 06:33 UT showing NOAA AR 11850. The green box represents the IRIS slit-jaw FOV at the start of the IRIS observations (08:09:43 UT). The blue box also represents the IRIS slit-jaw FOV but at the end of the IRIS observations (10:05:32 UT), as solar rotation compensation was not switched on. Finally the red box shows the CRISP FOV. North is up and the IRIS slit is positioned along the East-West direction.}
\vspace{0.5cm}
\label{context}
\end{figure}

The CRisp Imaging SPectropolarimeter \citep[CRISP;][]{2006A&A...447.1111S, 2008ApJ...689L..69S} is a dual etalon Fabry-P\'erot interferometer mounted in telecentric configuration at the SST. It takes high spatial-resolution images with a narrow passband in the spectral range between 500 and 860 nm. Spectral profiles can be constructed over the entire FOV by sequentially acquiring images over a given spectral window, 
typically at a rate of 4 spectral positions per second (or 1 spectral position in polarimetric mode). 
We obtained data in the \ion{H}{1}~656.3~nm (H$\alpha$) line, the \ion{Ca}{2}~854.2~nm line and the \ion{Fe}{1}~$630.2$~nm  line, the latter with polarimetry. 
H$\alpha$ was symmetrically sampled at 15 line positions, with equidistant steps of 20~pm, 
 \ion{Ca}{2}~854.2~nm was symmetrically sampled at 25 line positions, with equidistant steps of 10~pm, 
 and Stokes I, Q, U, and V filtergrams were acquired in the blue wing of \ion{Fe}{1}~$630.2$~nm at $-4.8$~pm. The temporal cadence of the CRISP data is 10.9~s, with a spatial sampling of $0.057 \arcsec$ per pixel. The SST was tracking solar rotation which resulted in a spatial overlap with the IRIS spectrograph slit decreasing from about 38\arcsec\ at the beginning to about 25\arcsec\ at the end of the sequence. The IRIS and SST FOVs are overlaid on a Hinode magnetogram in Figure~\ref{context}.
 
CRISP started observing at 08:44:26 UT, and continued until 10:29:13 UT, so that the temporal overlap with IRIS is 1~h 21~min. In this analysis we concentrate on the timespan with overlap between the IRIS and CRISP observations, i.e., 08:44:26 -- 10:05:32 UT, for a total of 420 IRIS rasters. 

The data were reduced using the CRISPRED pipeline \citep{2015A&A...573A..40D} which includes image reconstruction with the Multi-Object-Multi-Frame-Blind-Deconvoltution technique \citep[MOMFBD,][]{2005SoPh..228..191V}.  Residual seeing motions within one line scan were compensated with the method proposed by 
\citet{2012A&A...548A.114H} 
and large scale rubbersheet deformations due to differential seeing in the time series were removed using cross-correlation techniques \citep{1994ApJ...430..413S}. The seeing was good and the image quality benefited from the SST adaptive optics system 
\citep{2003SPIE.4853..370S}.

We aligned the CRISP data to the IRIS observations by scaling down to the IRIS pixel scale ($0.16 \arcsec$) and cross-correlating the far H$\alpha$ wing ($\Delta \lambda=-0.14$~nm) and IRIS \ion{Mg}{2}~h wing 283.2~nm slit-jaw images. Both channels show a very similar photospheric scene so that sub-IRIS pixel accuracy alignment was feasible. 

We have made extensive use of CRISPEX \citep{2012ApJ...750...22V}, a widget-based tool for effective analysis of multidimensional datasets. CRISPEX has been developed to include IRIS observations and is part of SolarSoft.

To investigate any coronal signal resulting from the small scale flux emergence events, we have also co-aligned simultaneous Solar Dynamics Observatory (SDO) AIA images \citep{2012SoPh..275...17L} from the 30.4~nm (chromosphere, transition region; $\log T=4.7$), 17.1~nm (upper transition region; $\log T=5.8$) and 19.3~nm (corona; $\log T=6.5$) channels.

\subsection{Line parameters and velocities of the bubble}
\label{line}

The magnetic properties of the emerging field are characterized using the linear polarization (LP) and the circular polarization (CP) degrees, which are proportional to the transverse and longitudinal components of the vector magnetic field respectively. In this case the \ion{Fe}{1} line was sampled at one point, thus we have single-wavelength \ion{Fe}{1}~$630.2$~nm Stokes I, Q, U and V maps. The mean linear and circular polarization degrees (see Paper I for a definition) get then reduced to ${\rm LP} = \sqrt{Q^2 + U^2} / I^2$ and ${\rm CP} = V/I$, respectively.

Regarding the ascent of the bubble through the different layers of the solar atmosphere, we have computed velocities of the emerging plasma in \ion{Ti}{2} 278.5 nm and \ion{Fe}{1} 630.25 nm (representing the photosphere), \ion{Ca}{2} 854.2 nm (representing the chromosphere) and \ion{Si}{4} 140.3 nm (representing the transition region). Since the photospheric \ion{Fe}{1} line was only sampled at a single wavelength point, we have not been able to compute this velocity in a standard way, but only using a proxy. This very simplistic proxy is based on the fact that a blueshifted profile will appear darker in the blue wing than a normal non-blueshifted profile. Therefore, plotting intensity at that single wavelength as a function of time will show darkenings that we interpret as upflows. We are aware that other mechanisms may lead to similar darkenings in the intensity maps.

To make the case stronger in the photosphere, we have selected one of IRIS photospheric lines, in particular the \ion{Ti}{2} 278.5 nm, which is formed at 220 km above $\tau_{500}=1$ \citep{2013ApJ...778..143P}. We have fitted a Gaussian to these absorption profiles in order to obtain the velocity.

\begin{figure}
\centering
\resizebox{\hsize}{!}{\includegraphics{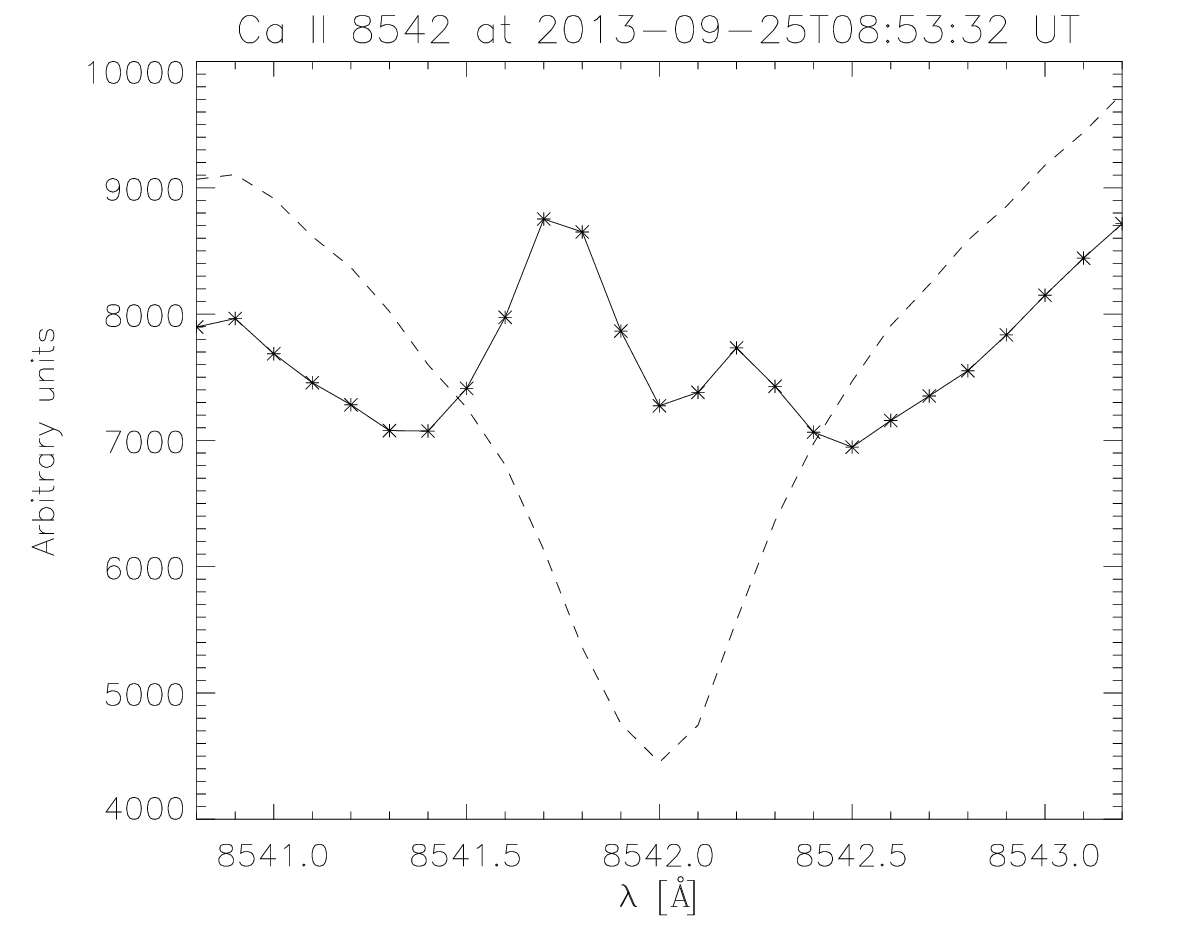}}
\caption{Example of \ion{Ca}{2} 854.2 nm profile inside the emerging plasma bubble (asterisks solid curve) obtained at 08:53 UT. A quiet Sun profile has been plotted for comparison (dashed line).}
\label{perf}
\vspace{0.5cm}
\end{figure}

The chromospheric velocity has been computed using another proxy. In the case of the \ion{Ca}{2} profiles, they present complicated characteristics within the emerging plasma bubble (e.g., double emission peaks), as shown in Figure~\ref{perf}.  This fact prevents us from using standard methods to calculate velocities, like bisectors, center-of-gravity or gaussian fits. In view of these complex profiles, we have chosen to apply a simple proxy to the LOS velocity, $(I_{blue} - I_{red})/(I_{blue}+I_{red})$, where $I_{red}$ and $I_{blue}$ correspond to intensities in the red (+700 m\AA) and blue (-700 m\AA) wings of the line respectively, far enough from the double emission peaks.


Finally, for the transition region velocities we have fitted a Gaussian to the \ion{Si}{4} emission profiles. 

None of the three Doppler velocities derived in the \ion{Fe}{1}, \ion{Ca}{2}, or \ion{Si}{4} lines admits a straightforward interpretation as a simple upflow, but they do allow us to follow the evolution, and especially the relative timings, of the rise of the magnetic bubble and the cold plasma contained within. 

\section{Results}
\label{results}

The field of view (FOV) shows a number of flux emergence events that evolve in a very similar manner: initially horizontal field appears within a granule. As the granule expands, two regions of vertical field (strong CP signal) and opposite polarities become evident at the edges. These regions are visible in the wing of the \ion{Ca}{2} 854.2~nm line as brightenings, while a dark bubble is formed in between the two vertical legs. In Paper~{\sc i} this set of events was interpreted to imply that a bubble of magnetic field was rising into the chromosphere. Here, IRIS observations will allow us to establish that the fate of this rising bubble depends on the topology of the pre-existing field above the flux emergence site.

In this section we consider two flux emergence events. After a global description, we will perform a more detailed analysis of the second event where, in addition to IRIS slit jaw images, we also have access to IRIS spectra.

\begin{figure*}
\centering
\resizebox{0.65\hsize}{!}{\includegraphics{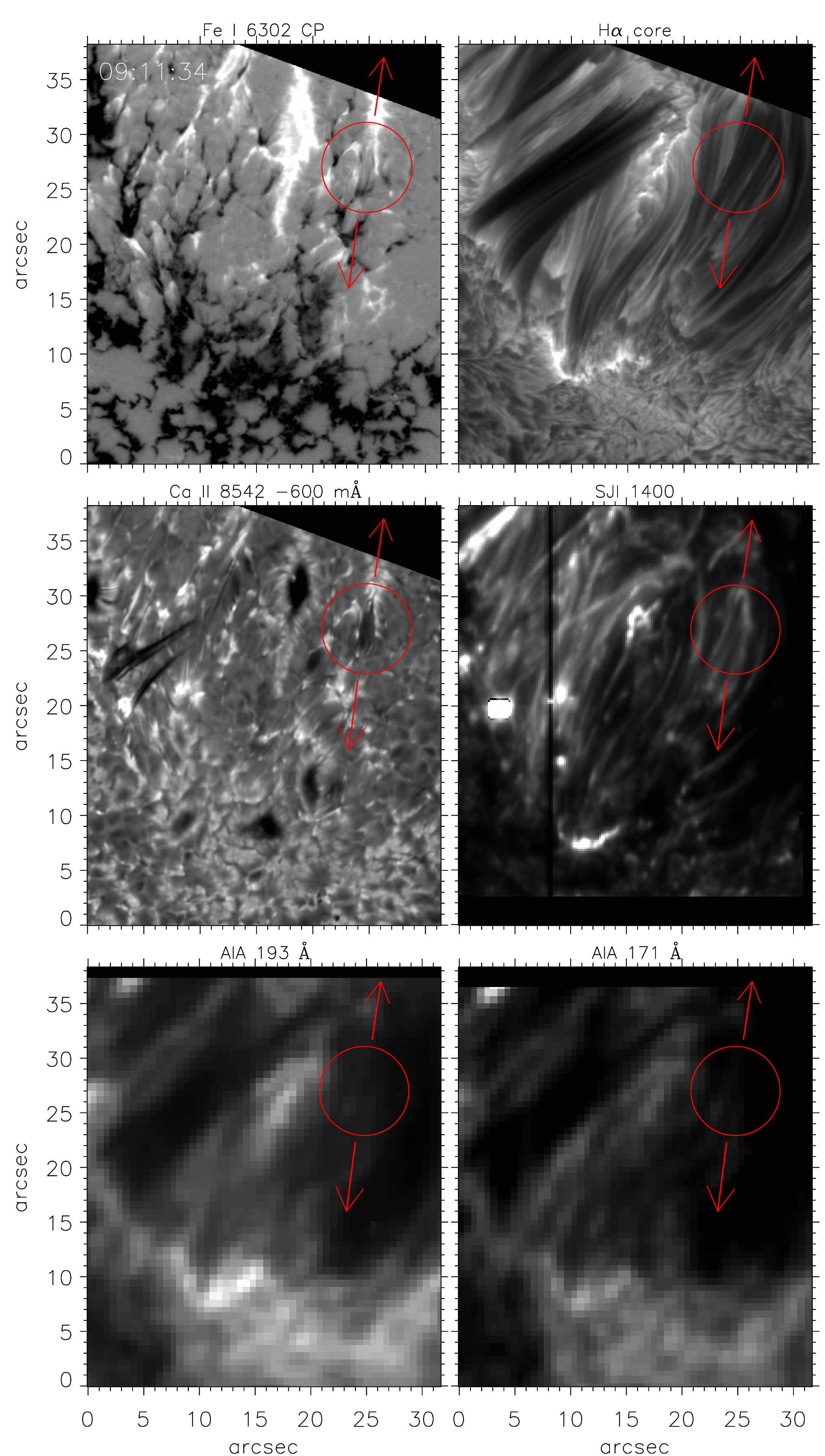}}
\caption{Overview of the FOV for case \#1 at 09:11:34~UT. Images have been rotated $90^{\circ}$ clockwise with respect to Figure~\ref{context}. The upper left panel shows a
photospheric magnetogram (circular polarization in the \ion{Fe}{1} line) presenting the FOV around the flux emergence event. The red circle marks the position of the
emerging flux in the six panels, while the red arrows show the direction of separation of the
opposite polarity magnetic legs with time. The upper right panel shows a CRISP filtergram taken at the core of the H$\alpha$
line. The mid left panel shows a \ion{Ca}{2} 854.2 $-0.06$ ~nm filtergram. The mid right panel presents a 140 nm IRIS slit-jaw image. The slit-jaw image shows
the position of the IRIS slit, now as a dark vertical line. Both the H$\alpha$ fibrils and the 140 nm loops trace the presence of field lines. The lower panels present AIA images at 19.3 and
17.1 ~nm, respectively. Contamination from the transition region in these coronal lines seems to be limited.}
\label{caso1_mosaic}
\end{figure*}

\subsection{Tomographic view from the photosphere to the corona}

Figure~\ref{caso1_mosaic} shows the FOV covered by both SST and IRIS observations from which we have selected two flux emergence events, in the following named case \#1 and case \#2. The upper left panel shows a map of circular polarization (CP) in \ion{Fe}{1}. We find two strong bands of oppositely directed vertical field. Flux emergence takes place around the polarity inversion line (PIL) located between $[x,y]=[3\arcsec,33\arcsec]$ and $[28\arcsec,15\arcsec]$, visible as small pairs of opposite polarity feet in the \ion{Fe}{1}~630.2 CP image. The upper right panel is a CRISP filtergram in the H$\alpha$ core. A large number of H$\alpha$ fibrils connect the two main polarity bands, oriented in the SE -- NW direction but with varying angles, sometimes one set of fibrils covering another. 
The two regions of brightest emission in the H$\alpha$ core correspond to dark pore-like regions in the \ion{Ca}{2} 854.2 $-0.06$ ~nm filtergram (central left panel). These regions are also bright in the IRIS 140~nm slit jaw image (central right panel), and have emission in both the 17.1~nm and 19.3~nm AIA channels displayed in the bottom two panels, implying that they are regions connected to transition region and low lying coronal temperatures. 
The entire SE portion of the FOV is formed by unipolar plage, and looks moss-like in the AIA channels, with no long fibrils in the H$\alpha$ core, but rather filled with short dynamic fibrils. The darkest fibrils in H$\alpha$ are also quite dark in both AIA channels, indicative of cool neutral hydrogen and helium bound-free absorption. The loops visible in the IRIS SJI image are oriented much as the H$\alpha$ fibrils, but there do not seem to be any transition region loops corresponding to the darkest and highest \citep{2012ApJ...749..136L} H$\alpha$ fibrils joining $[x,y]=[5\arcsec,17\arcsec]$ and $[15\arcsec,32\arcsec]$. There are also a number of 
bright points in the IRIS 1400 SJI image that do not obviously correspond to features in the other filtergrams. 

\subsection{Case \#1}

Emergence event case \#1 arises within the positive polarity band. Its positition is highlighted in Figure~\ref{caso1_mosaic} by a red circle, and the arrows show the direction of separation of the magnetic legs. The snapshot was taken at 09:11:34~UT. We can clearly see a positive polarity leg (white) and a negative polarity leg (black) within the red circle in the CP map. They seem to have a half-moon shape. The H$\alpha$ core image shows several cold loops lying above event \#1 (here we are assuming that the H$\alpha$ core is formed above the flux emergence event, see \citet{2012ApJ...749..136L}; we will discuss this further in Section~\ref{disc}). The newly emerged field rises under these cold loops, which eventually may hinder its ascent. The \ion{Ca}{2} 854.2 $-0.06$ ~nm filtergram shows the magnetic legs as brightenings (particularly the positive polarity leg), while the space between them appears as a dark patch (i.e., the dark bubble). Similarly to the core of the H$\alpha$ line, the 140~nm IRIS slit-jaw image gives hints of the configuration of the overlying loops, but it also shows the TR response to the emergence event. At the time the coronal images were obtained there is no obvious sign of flux emergence. More details of the rising plasma evolution will follow in the next paragraphs.

\begin{figure*}
\centering
\resizebox{\hsize}{!}{\includegraphics{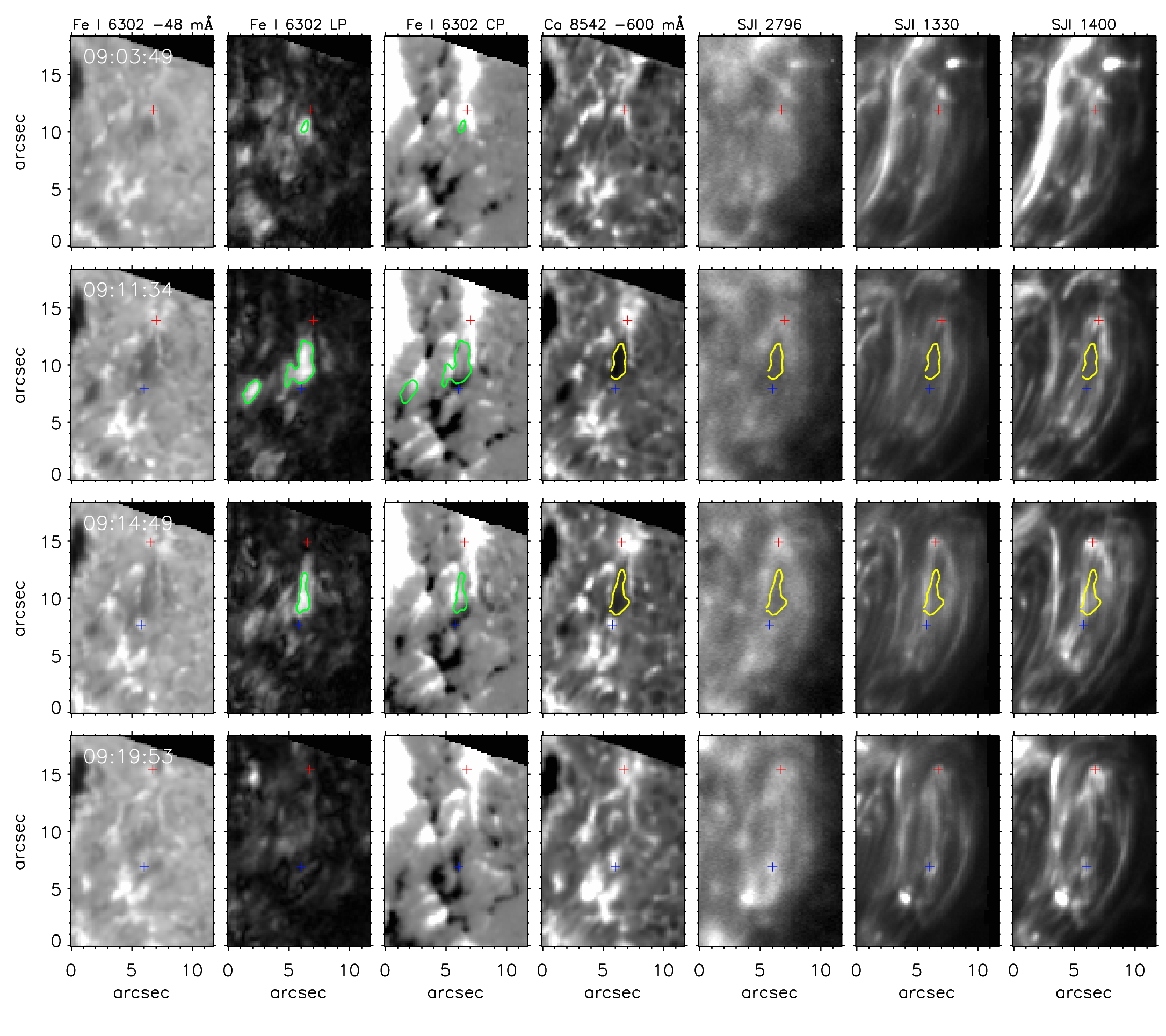}}
\caption{Temporal evolution of several observables corresponding to flux emergence case \#1.  From left to right:
  line wing intensity  at 630.25-0.0048~nm, photospheric LP and CP maps, a filtergram at \ion{Ca}{2} 854.2 $-0.06$
  ~nm, and three of the IRIS slit-jaw images (279.6, 133 and 140 nm respectively). The dark bubble appears clearly
  in some of the \ion{Ca}{2} 854.2 $-0.06$ ~nm filtergrams, in particular at 09:11:34 and 09:14:49~UT. Its
  contours have been highlighted in yellow. The green contours overplotted on the LP and CP maps indicate linear
  polarization signals of 4\% of the intensity. The red and blue crosses mark the magnetic feet of
  opposite polarities (red for positive polarity and blue for negative polarity). Note that the negative
  polarity only appears in the second time-scan, while the positive leg was already there from the beginning of
  the time series. As time goes by the two legs of opposite polarities separate from each other. The time stamps
  in the first column corresponds to SST observations. Loops connecting both polarities are seen in all slit-jaw
  images at 09:14:49 and 09:19:53~UT.}
\label{case1_sstiris}
\end{figure*}

Figure~\ref{case1_sstiris} and the accompanying movie summarize the time evolution of event \#1. From left to right the figure shows seven observables: line wing intensity at \ion{Fe}{1} 630.25-4.8~pm, photospheric LP and CP maps, a filtergram at \ion{Ca}{2} 854.2 $-0.06$ ~nm, and three of the IRIS slit-jaw images (279.6, 133 and 140~nm respectively). Four particular instants in the life of this flux emergence event are highlighted. The movie shows the same observables, plus a number of filtergrams scanning the \ion{Ca}{2} 854.2~nm and H$\alpha$ lines, as well as AIA/SDO filtergrams for completeness.

The first glimpse of increased linear polarisation appears at 09:01:07~UT in the LP map, south of a small patch of pre-existing positive
polarity ($[x,y]=[5\arcsec,10\arcsec]$). From this timestep on, the LP signal grows both in size and in strength until reaching a maximum
length of about 7\arcsec\, at 09:18:37~UT. The positive polarity is already present at the beginning of the time sequence (08:59:01~UT). This patch of positive
polarity drifts towards the north, merging with another positive patch ($[x,y]=[6.5\arcsec,11\arcsec]$) at 09:01:50. At
09:01:39~UT a slight darkening appears in the \ion{Fe}{1} 630.25-4.8~pm intensity below the brightened positive pole, coinciding spatially
with the LP patch. This dimming in the wing of the \ion{Fe}{1} line itself suggests the presence of a blueshift in the photospheric
layers. The positive polarity also shows a brightening in the \ion{Ca}{2} 854.2 $-0.06$~nm filtergram. At 09:03:38~UT the dark bubble appears in that same filtergram. The first row of Figure~\ref{case1_sstiris} shows the situation at 09:03:49~UT, with dimming in \ion{Fe}{1}~$630.2$~nm, an evident LP patch, CP showing mainly pre-existing polarities, and a dark bubble in \ion{Ca}{2} 854.2 that has just appeared at $-0.06$~nm. At this time no related structure is visible in the IRIS slit-jaw images.

Just one minute later, at 09:04:43~UT, there is a clear brightening of the loops in the 140~nm IRIS slit-jaw image (slightly dimmer in 133~nm)
that appears to be associated with the pre-existing configuration, and the \ion{Mg}{2} 279.6 nm slit-jaw image shows intensity enhancements near the location of the positive polarity. One minute later the dark bubble is observed in the \ion{Ca}{2} 854.2 $\pm 0.03$~nm filtergrams. The first glimmer of negative polarity is finally seen 
in the CP map at 09:05:58~UT. This newly emerged negative foot appears at $[x,y]=[5\arcsec,7.5\arcsec]$ in between pre-existing, $V$-shaped, bigger positive and negative patches. Later on, presumably as the magnetic bubble expands and reaches the mid-to-high chromosphere, a darkening becomes visible in the \ion{Mg}{2} 279.6~nm image (09:08:30~UT) in a region that was bright before. The new negative polarity footpoint brightens first in the \ion{Ca}{2} 854.2 $-0.06$~nm filtergram at 09:09:45~UT and then, with a three minute delay (09:12:50), in the \ion{Mg}{2} 279.6 nm slit-jaw image. The second row of Figure~\ref{case1_sstiris} portrays these events. All IRIS images show at this point a brightening coinciding with the positive polarity.

At 09:13:05~UT a new loop-like feature -- at a different angle with respect to the older loops -- lights up connecting the recently emerged positive and negative polarities. The brightening of the loops observed in IRIS is highly variable and the new loops disappear rapidly from view. Another similar loop lights up and is portrayed in the third row of Figure~\ref{case1_sstiris} at 09:14:49~UT. At this time, thirteen minutes after the appearance of the LP patch in the photosphere, the newly emerged magnetic polarities show corresponding brightenings in all wavelengths, from the photospheric \ion{Fe}{1} line to transition region IRIS slit-jaws 133 and 140~nm, while the dimming produced by the dark bubble between the polarities is visible in most intensity images.

Assuming that the transition layer is situated at most 5000~km above the photosphere, we find that 
a delay of around thirteen minutes between the emergence of the LP patch into the photosphere and the loop-like connection joining our polarities, gives an average upward speed for the emerging plasma of 6.4 km\,s$^{-1}$. This speed is consistent with the observations and simulations presented in Paper~{\sc i}, which reported average rising speeds up to the mid-chromosphere of 5.2 km\,s$^{-1}$. A horizontal expansion speed of 4.5 km\,s$^{-1}$, as we observe here, is also in agreement with that obtained in Paper~{\sc i}.

At 09:16:22~UT the negative polarity ``sparkles'' in the slit-jaw images at 133 and 140~nm while it drifts towards the larger, pre-existing, southern $V$-shaped negative polarity. It is remarkable that, between 09:15:55 and 09:18:07~UT, all AIA images show a large loop similar to the one seen in H$\alpha$ core, H$\alpha$ $-0.04$ ~nm and the slit-jaw images (see movie). This large structure probably corresponds to long pre-existing loops rather than loops formed by the emergence event we are describing.

One and a half minutes later, at 09:17:54~UT, the movie shows that the dark bubble has disappeared from the \ion{Ca}{2} 854.2 $-0.06$~nm and \ion{Mg}{2} slit jaw images. Remarkably the dark bubble is, at this time, very well visible in the core of H$\alpha$.

The fourth row of Figure~\ref{case1_sstiris} portrays a situation in which the footpoints of the loop are bright at all wavelengths, there is a hint of a magnetic loop connecting both polarities in IRIS images, there are no traces of the dark bubble at any wavelength, and there is no significant remaining LP signal. Later on, between 09:18:31 and 09:25:31~UT, the movie shows that the AIA wavelengths present brightenings coinciding spatially with our positive polarity. 


\begin{figure}
\centering
\resizebox{0.75\hsize}{!}{\includegraphics{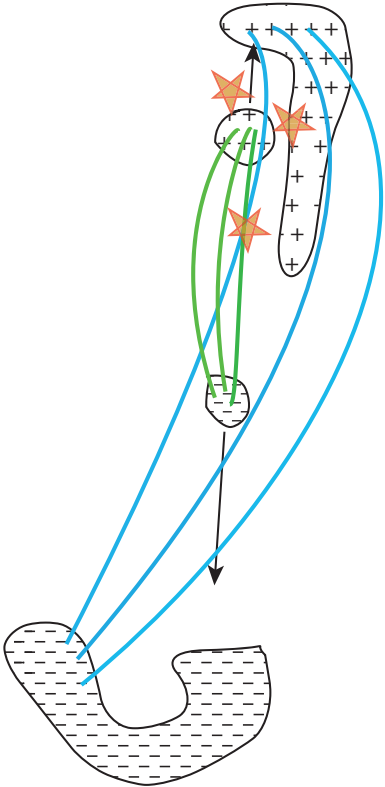}}
\caption{Schematic interpretation of flux emergence event \#1. The cartoon represents both the pre-existing magnetic field lines (blue curves) and the smaller emerging loop (green curves). The new flux pierces the photosphere close to the pre-existing positive polarity. Then, the legs of opposite polarities drift away from each other, as is typical in granular-sized flux emergence events (black arrows). The low-lying newly
  emerged flux forms loops that rise into the chromosphere, transition region and corona at an angle to
  pre-existing loops. Thus, large field gradients in the vicinity of the positive polarity foot points arise and may
  lead to reconnection driven heating (example red stars), eventually causing irregular transition region emission along both
  pre-existing and newly emerged loops.}
\label{fig:cartoon_case1}
\end{figure}

The events recorded here lead to an interpretation of case \#1 as
illustrated in Figure~\ref{fig:cartoon_case1}: Initially horizontal field emerges through the photosphere 
in the vicinity of the $\it V$-shaped positive polarity forming one of the pre-existing loop footpoints, while the negative 
associated polarity is some 10$\arcsec$ towards the south. While short, the loops stretching between these polarities span several granules. The emerging field rises through the chromosphere at an angle to the pre-existing banana-shaped loops. The location where these field systems first interact, near the positive polarity, is probably the site of 
reconnection and associated currents causing heating and thus irregular brightenings all along the field connecting the footpoints.

\begin{figure*}
\centering
\resizebox{\hsize}{!}{\includegraphics{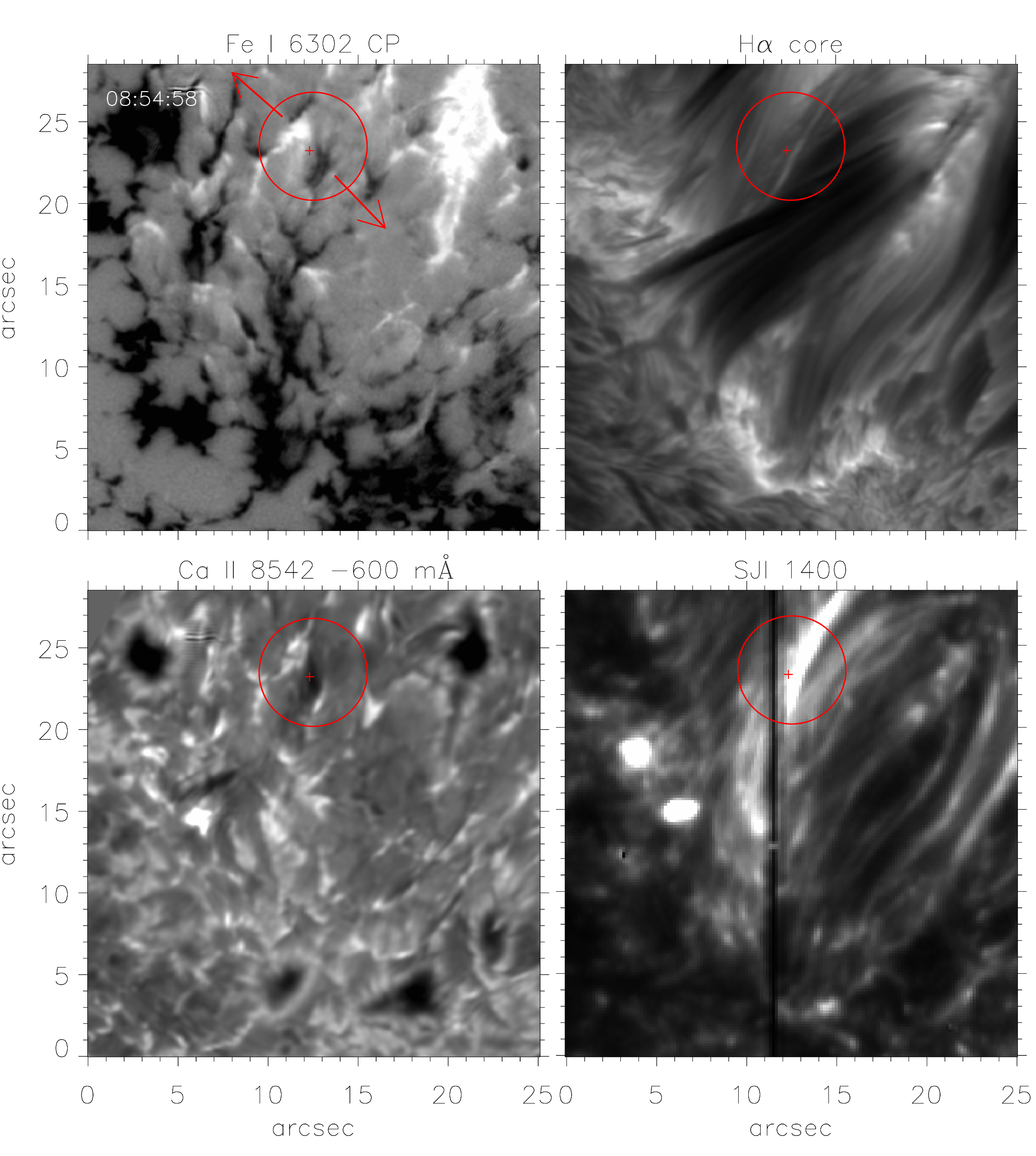}}
\caption{Four snapshots showing the magnetic topology (traced by the loops and fibrils overlying the flux mergence event) for case \#2 at 08:54:58~UT. The upper left panel shows a photospheric magnetogram of the FOV that surrounds the flux emergence event. The red circles mark the location of
the flux emergence event in the four panels, while the red arrows show the direction of separation of the
opposite polarity legs with time. The cross identifies a pixel at the center of the dark bubble. This pixel will
be referred to in coming figures. The upper right panel shows a CRISP filtergram taken at the core of the
H$\alpha$ line. The lower left panel shows a \ion{Ca}{2} 854.2 $-0.06$ ~nm filtergram. Finally, the lower right panel presents a 140 nm IRIS slit-jaw image. This image has been saturated for clarity.}
\label{case3_mosaic}
\end{figure*}

It is worth noting that the AIA images seem to respond to some of the loops and brightenings seen in the core of the H$\alpha$ line and, sometimes, in the IRIS slit-jaw images. Those loops and brightenings are most likely related to the pre-existing activity. We have not found a {\it clear} imprint of flux emergence case \#1 in the AIA channels corresponding to \ion{He}{2} 30.4~nm (transition region), or \ion{Fe}{9} 17.1~nm and \ion{Fe}{10} 19.3~nm (transition region and coronal). We have examined the movie for another 20~minutes in order to detect obvious brightenings connected with our flux emergence, to no avail. Thus, these fields either do not reach the temperatures covered by AIA images, or need more than 33 minutes to reach those layers since they appear in the photosphere.

\subsection{Case \#2}

Case \#2 occurs in the same FOV as case \#1 but at an earlier time, 16$\arcsec$ to the east of case \#1. This particular event occurs just at the position of the IRIS slit (shown as a black vertical line in the 140~nm IRIS slit-jaw image in Figure~\ref{caso1_mosaic} and in Figure~\ref{case3_mosaic}). Therefore, this case will allow us to derive spectral information and line profiles (including LOS velocities of the upflowing plasma) from several spectral lines formed in the upper chromosphere and the transition region.

Figure~\ref{case3_mosaic} presents a close-up of the region where the event occurs. Again, the purpose of this mosaic of images is to show the general configuration of the magnetic topology at a specific time. Case~\#2 is found east of a set of very dark fibrils lying at an angle to, and above, the fibrils joining the main polarities. The location of this particular event is marked with a red circle in all four images. Note that the orientation of the emerging field in this case is close to $90 ^\circ$ compared to the overlying ambient fields (seen as H$\alpha$ fibrils), a much larger relative angle than that found in case \#1. The red cross highlights a pixel within the dark bubble which will be the object of attention in forthcoming figures. 

The upper left panel is a map of CP in \ion{Fe}{1} in which the positive (white) and negative (black) magnetic legs are easily seen. The FOV is dominated by magnetic fields of negative polarity. The upper right panel shows a simultaneous CRISP filtergram in the H$\alpha$ core. As for case \#1, the diagonally oriented dark fibrils lie above the flux emergence event which is unfolding underneath. The lower left panel is a  \ion{Ca}{2} 854.2 $-0.06$ ~nm filtergram. Again, the magnetic legs are seen as brightenings at this wavelength, while a dark bubble forms in between the two opposite polarities. The lower right panel shows an IRIS 140~nm slit-jaw image as well as the position of the spectrograph slit. \ion{Si}{4} loops lying above our flux event are apparent at these transition region temperatures. These loops are oriented closer to the N-S direction and correspond to the apparently lower lying H$\alpha$ fibrils rather than to the overlying SE-NW oriented darker fibrils. 

\begin{figure*}
\centering
\resizebox{0.8\hsize}{!}{\includegraphics{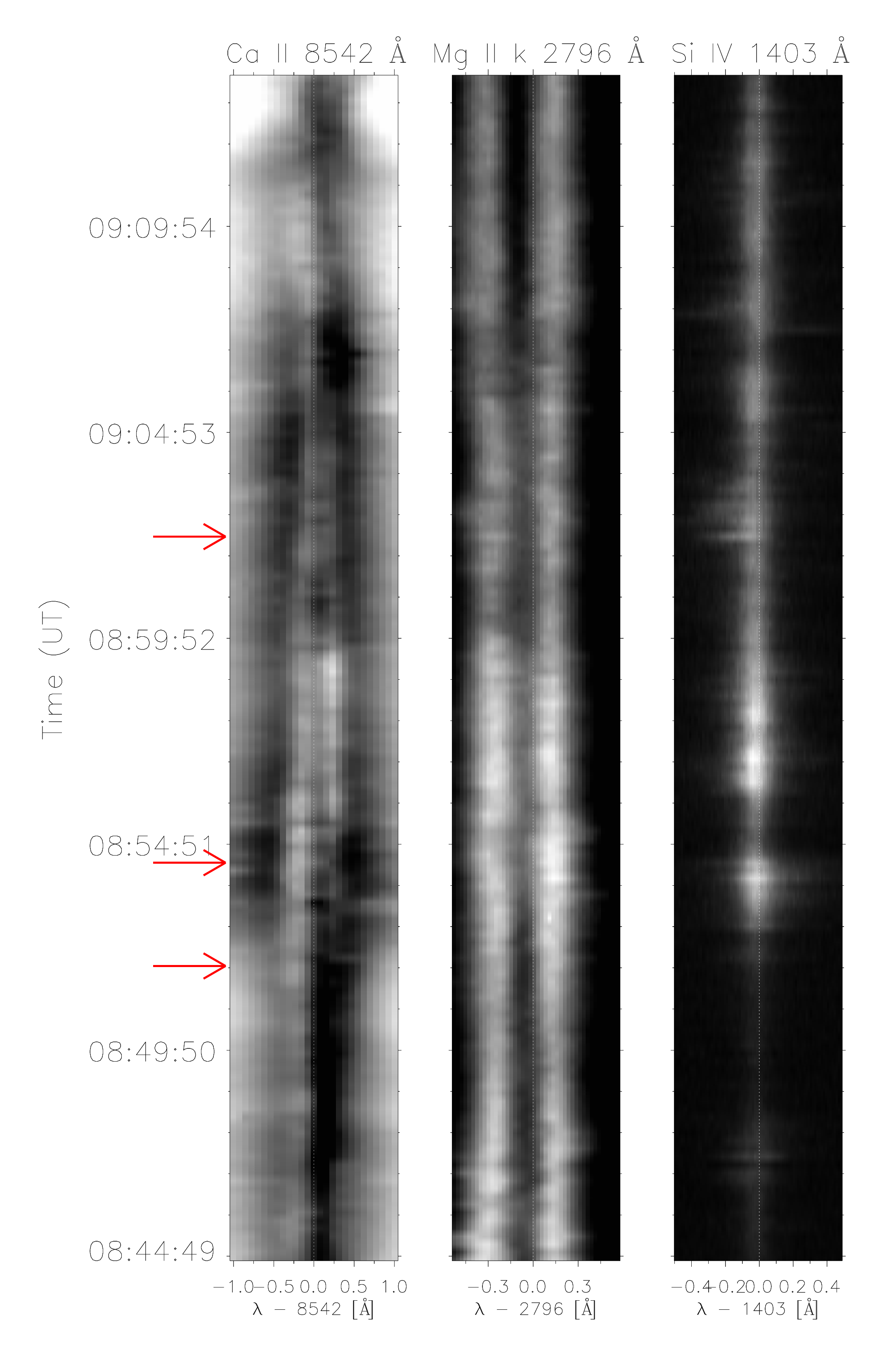}}
\caption{Time-sliced spectra at the pixel marked by a cross in Figure~\ref{case3_mosaic} for three lines,
\ion{Ca}{2} 854.2 nm, \ion{Mg}{2} k 279.6 nm and \ion{Si}{4} 140.3 nm. The time span represented here covers the whole duration of the
flux emergence event. The three red arrows mark specific moments in the evolution of case \#2 (see
text).}
\label{case3_spectra_lt}
\end{figure*}

In this particular case IRIS provided not only images but also spectra in the far and near ultraviolet, of which we have analized the \ion{Si}{4} 140.3 and the \ion{Mg}{2} k 279.6 nm lines. We use these data to compute upward velocities and time delays for the passage of the magnetic bubble through the different atmospheric layers. Figure~\ref{case3_spectra_lt} shows time-sliced spectra for three different lines: \ion{Ca}{2} 854.2 nm (taken by CRISP) and \ion{Mg}{2} k 279.6 and \ion{Si}{4} 140.3 nm (taken by IRIS), in the pixel highlighted in Figure~\ref{case3_mosaic}. These lines provide information of the lower-to-mid chromosphere, upper chromosphere, and transition region respectively. The time span chosen in this figure covers the whole duration of event \#2, which lasts 20 minutes approximately from 08:50 till 09:10 UT.

The red arrows in Figure~\ref{case3_spectra_lt} mark three particularly interesting moments in the temporal evolution of case \#2. The lower arrow at 08:51:44~UT shows the moment in which the LP signal became significant, indicating the emergence of horizontal magnetic fields into the photosphere. The middle arrow, at 08:54:16~UT, indicates a blueshift in the \ion{Ca}{2} 854.2 nm line which signals the passage of the bubble through the lower chromosphere. The \ion{Mg}{2} k 279.6 line (mid panel) also shows a moderate blueshift around 09:00:50~UT, which occurs temporally between the \ion{Ca}{2} and the \ion{Si}{4} blueshifts. Finally, the upper arrow marks the time of the blueshifts in the \ion{Si}{4} 140.3 nm line (reaching up to 40 km\,s$^{-1}$), at 09:02:11~UT, marking the arrival of the bubble to the transition region and the interaction with the pre-existing field there. 

\begin{figure}
\centering
\resizebox{\hsize}{!}{\includegraphics{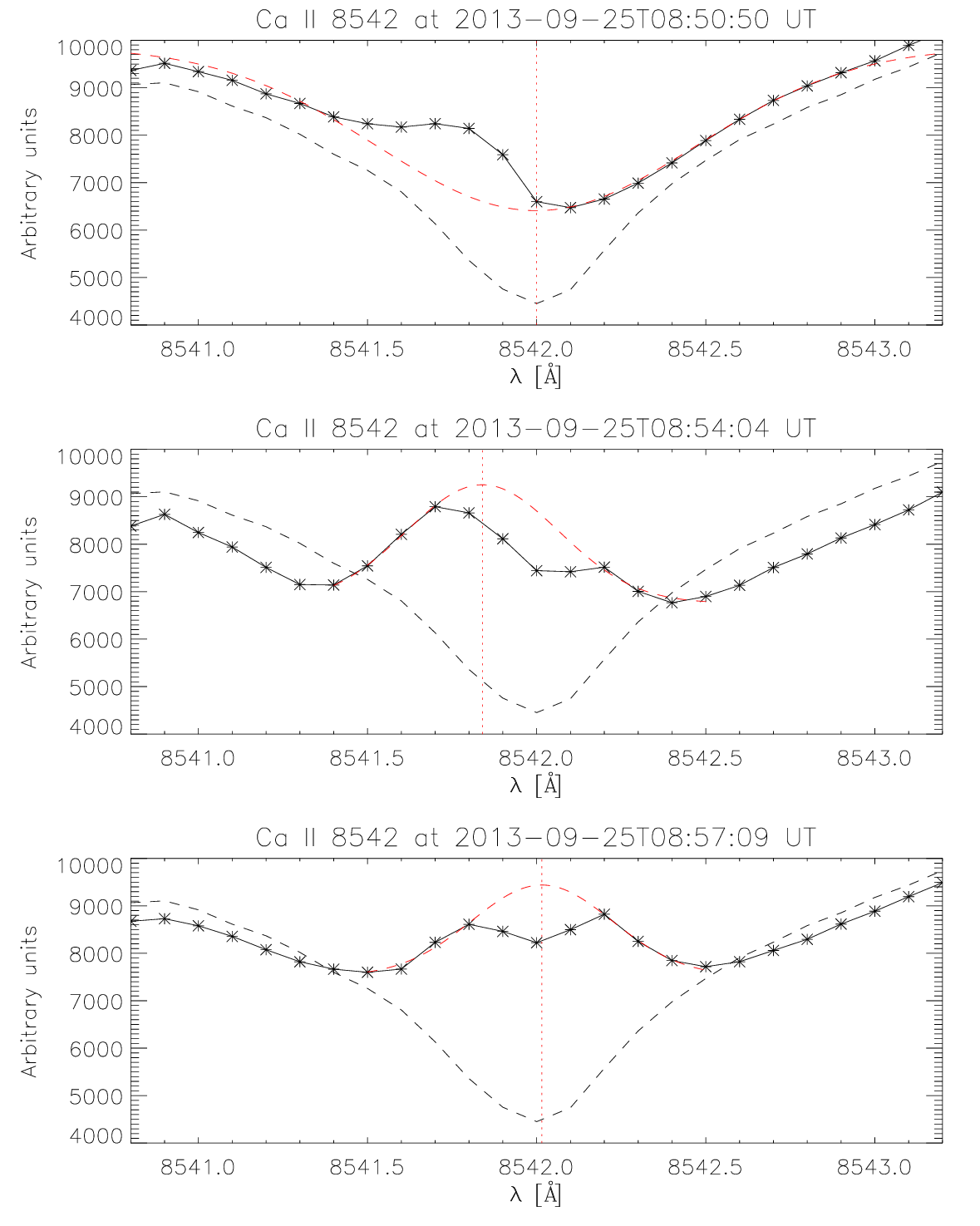}}
\caption{Three cuts in the \ion{Ca}{2} panel of Figure~\ref{case3_spectra_lt}, showing the details of three \ion{Ca}{2} 854.2 nm profiles, both before, during, and after the passage of the magnetic bubble through the chromosphere (solid black lines). A quiet Sun profile has been plotted for comparison on each panel (black dashed line). The red dashed curves represent a Gaussian fit to each profile, whose center is noted with a red dotted vertical line.}
\label{referee1}
\end{figure}

\begin{figure*}
\centering
\resizebox{0.8\hsize}{!}{\includegraphics{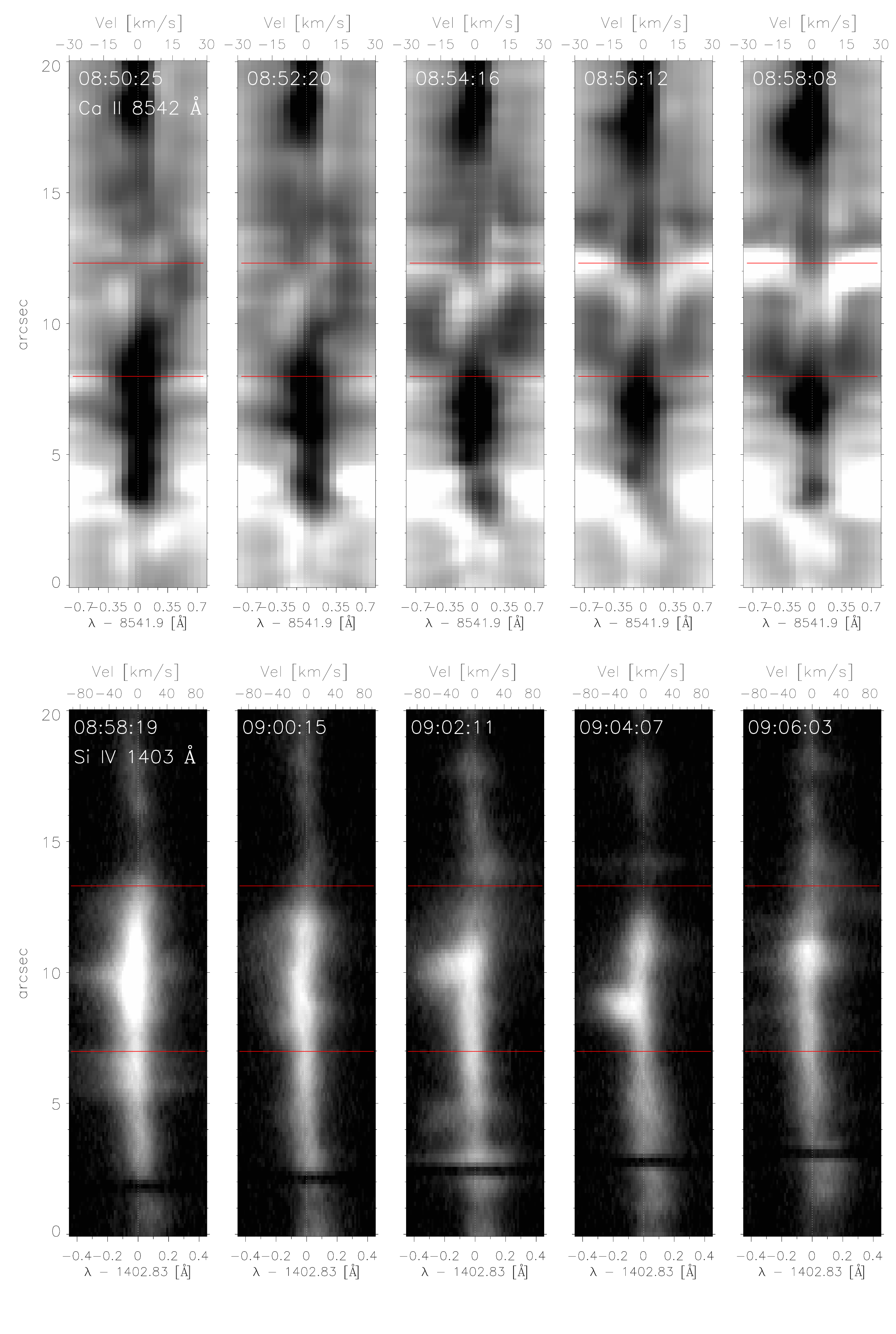}}
\caption{Spectra of the \ion{Ca}{2} 854.2 nm line (upper panels) and the \ion{Si}{4} 140.3 line (lower panels) centered at
the pixel marked by a cross in Figure~\ref{case3_mosaic} $\pm$ 10$\arcsec$. The IRIS slit is oriented in the same direction as in Figures~\ref{caso1_mosaic} and \ref{case3_mosaic}. Notice
the blueshift in \ion{Ca}{2} 854.2 nm at 08:54:16~UT (mid upper panel, corresponding to the middle arrow in Figure~\ref{case3_spectra_lt}). Another prominent blueshift, this time in \ion{Si}{4} 140.3 (mid lower panel,
corresponding to the upper arrow in Figure~\ref{case3_spectra_lt}), can be easily observed at 09:02:11~UT,
presenting a delay of about eight minutes with respect to the chromospheric blueshift. Some other spectra at nearby
timesteps have been included for comparison (both before and after the mentioned blueshifts), showing that these
blueshifts are of a transient nature. The white dotted vertical lines show the line centers for reference. The red horizontal lines mark the perimeter of the dark bubble, and coincide with the upper and lower bright legs. The drifting dark band at around x=2\arcsec\, is the fiducial mark on the IRIS spectrograph slit.}
\label{case3_spectra_ly}
\end{figure*}

Figure~\ref{referee1} shows the detailed shape of three \ion{Ca}{2} 854.2 nm profiles at three different times: before (08:50:50 UT), during (08:54:04 UT), and after (08:57:09 UT) the passage of the magnetic bubble through the chromosphere. These profiles are horizontal cuts in the \ion{Ca}{2} panel of Figure~\ref{case3_spectra_lt} at the mentioned times. In addition we have overplotted a Gaussian fit to each profile, and only the middle panel, at 08:54:04 UT, shows a blueshift of -6 km\,s$^{-1}$ marking the passage of the upflowing plasma. The purpose of this figure is to make it easier to identify the \ion{Ca}{2} blueshifts of Figures~\ref{case3_spectra_lt} and \ref{case3_spectra_ly} and to give a value to this chromospheric upflow.

Figure~\ref{case3_spectra_ly} provides further supporting evidence for 1) the fact that the mentioned blueshifts happen {\it within} the magnetic dark bubble area, 2) the existence of blueshifts in the chromosphere and in the transition region with an eight minute delay in between, and 3) the fact that these blueshifts are of a transient nature. The upper panels of the figure show the \ion{Ca}{2} 854.2~nm spectra from the pixel marked by a cross in Figure~\ref{case3_mosaic} $\pm 10 \arcsec$ at five different time steps. The chromospheric plasma seems to move up intermittently, since the \ion{Ca}{2} 854.2~nm spectra shown in Figure~\ref{case3_spectra_lt} present a burst-like evolution between 08:53:00 and 08:56:00 UT. The lower panels show \ion{Si}{4} 140.3~nm spectra in the same pixel, again at five different time steps. Notice the blueshift in \ion{Ca}{2} at 08:54:16~UT (mid upper panel, which corresponds to the middle arrow in Figure~\ref{case3_spectra_lt}). Another prominent blueshift, this time in \ion{Si}{4} (mid lower panel, corresponding to the upper arrow in Figure~\ref{case3_spectra_lt}), is observed at 09:02:11~UT presenting an eight minute delay with respect to the chromospheric blueshift. In order to prove that the observed blueshifts are of a transient nature (that is, they only exist during the passage of the bubble on its way up as opposed to commonly occurring) we have included spectra at nearby timesteps -- both before and after the highlighted blueshifts -- for comparison. Those nearby spectra show either no significant blueshifts, or if they do (like the \ion{Si}{4} spectra at 08:58:19 and 09:04:07~UT) they are also short-lived. The red horizontal lines mark the perimeter of the dark bubble (which increases its size with time as described in Paper~{\sc i}). Note that {\it all} the mentioned blueshifts happen within the area enclosed by the dark bubble.

\begin{figure*}
\centering
\resizebox{\hsize}{!}{\includegraphics{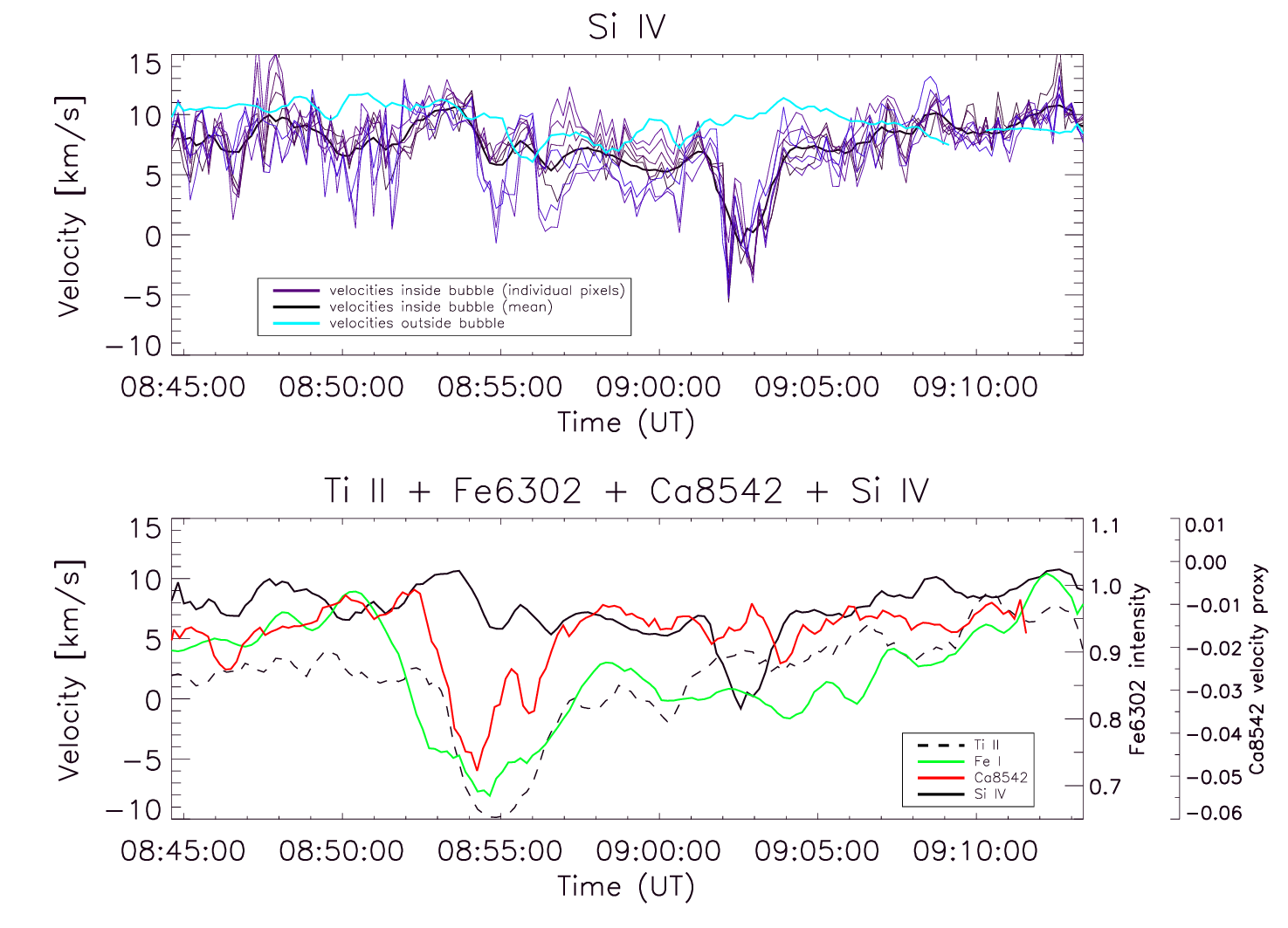}}
\caption{Upper panel: \ion{Si}{4} velocities as a function of time for several pixels within the dark bubble (purple lines) and their
smoothed mean (black line). Note the bump produced in the velocities at around 09:03:00~UT, marking the existence of an upflow of
more than 8 km/s in the transition region. Compare to the averaged velocities of pixels outside of the flux emergence region (cyan line). Lower
panel: velocities measured in the photosphere (green line and dashed black line), low chromosphere (red line) and transition region (solid black
line) as a function of time. Note the successive bumps happening with a certain delay in the different lines, proving how the upflowing plasma traverses different layers in the solar atmosphere. The \ion{Ti}{2} velocity has been multiplied by 10 for display purposes.}
\label{velos}
\end{figure*}

Finally, Figure~\ref{velos} gives an estimate of the order of magnitude of the upflows and also of the delays between them. The upper panel shows  \ion{Si}{4} velocities as a function of time, for eight pixels within the dark bubble (purple lines). The spatial average of those pixels has been smoothed in time using a 5-step running average. This smoothed mean is represented by the black line. For comparison, an average of pixels outside of the flux emergence region is plotted as a cyan line. Notice the bump produced in the velocities at around 09:03:00~UT, which is only detected by the pixels belonging to the bubble but not by the pixels outside the magnetic bubble. The bump corresponds to upflows of about 8 km\,s$^{-1}$ in the transition region. Notice also how some individual pixels within the bubble present transient upflows at other times (e.g. at 08:55:00 UT) that are not captured by the smoothed mean. It is important to realise that the 8 km\,s$^{-1}$ upflows found represent the result of a single Gaussian fit to the \ion{Si}{4} spectra. Double Gaussian fits show that the \ion{Si}{4} line is split into two components during the event, as is also evident from Figure~\ref{case3_spectra_lt} and Figure~\ref{case3_spectra_ly}. The more blueshifted component shows velocities of 40 to 50 km\,s$^{-1}$ in bursts, presumably due to reconnection between the emerging bubble and the ambient field (indicating that the bubble has reached the dark appearing canopy of H$\alpha$ fibrils), while the other component is blueshifted by at most a few km\,s$^{-1}$. The Doppler shift derived from the single gaussian fit of Figure~\ref{velos} is therefore showing the combined effect of reconnection bursts and the rising bubble, and cannot be considered a real velocity but rather a time marker.

The lower panel plots velocities and velocity proxies measured in the photosphere (green and dashed black line), low chromosphere (red line) and transition region (black line) as a function of time. In principle we would expect that the \ion{Fe}{1} intensity darkening and the \ion{Ti}{2} velocity upflow, seen at about 08:55 UT, would occur simultaneously, since both sample the photosphere at similar heights. However, there is a slight delay between the green and the dashed line which we believe can be explained from geometrical considerations. The SST \ion{Fe}{1} intensities can be measured at any point within the dark bubble, and in particular we have chosen its center. On the other hand, the IRIS \ion{Ti}{2} velocity can only be measured at the slit positions, which roughly coincide with the bubble edge. In fact, Figure~\ref{case3_mosaic} shows that none of the slits positions coincide with the bubble center, marked by the red cross. As the bubble rises through the photosphere, presumably shaped as a semisphere (see Paper I), the center will be detected first, while the velocity near the edges will take longer to be measured. This geometry also explains why the \ion{Ti}{2} velocity is smaller than the other velocities, since the expansion near the bubble edge is at an angle with respect to the line of sight.  

The successive bumps both in velocity and intensity as a function of time prove how the upflowing plasma traverses different layers in the solar atmosphere. Nine minutes separate the protrusion through the photosphere from the arrival at the transition region.

\subsection{Hot explosions above case \#2}

\begin{figure}
\centering
\resizebox{\hsize}{!}{\includegraphics{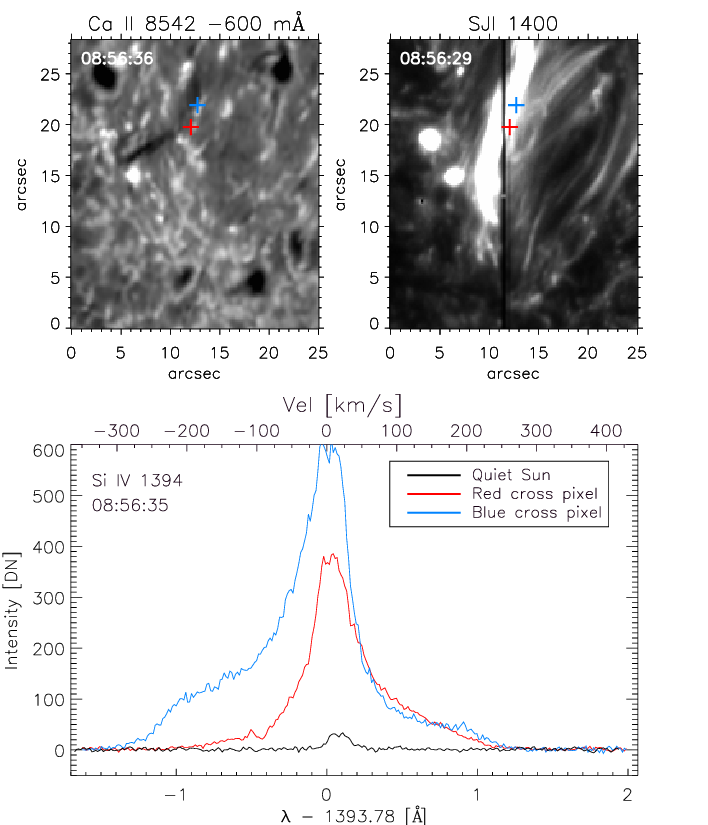}}
\caption{Upper panels: (left)  \ion{Ca}{2} 854.2 $-0.06$~nm filtergram showing the dark bubble and its bright magnetic feet just north of the red and blue crosses, indicative of newly emerging flux; (right) \ion{Si}{4} 140.3~nm slit-jaw image taken at 08:56:29~UT showing a set of highly time-variable loops that overlie the emerging flux. Lower panels: \ion{Si}{4} 139.3~nm spectra shown for the two locations marked by the blue and red crosses in the upper panels. In addition, a `quiet' \ion{Si}{4} 139.3~nm spectrum (black curve) is shown for comparison. Note the extremely large Doppler shifts (blue curve) and the absorption dent at -100~km$\,$s$^{-1}$ (red curve) seen in the blue wing of the \ion{Si}{4} 139.76~nm line, presumably caused by the existence of \ion{Ni}{2} 139.33 nm \citep[see][]{2014Sci...346C.315P}.}
\label{fig:reconnect_si4}
\vspace{0.5cm}
\end{figure}

The region around flux emergence case \#2 shows quite interesting transition region dynamics before, during and after the events described above. These dynamics are connected with the lower set of overarching fibrils and loops visible in the H$\alpha$ core and \ion{Si}{4} slit-jaw images in Figure~\ref{case3_mosaic}: the H$\alpha$ core image shows that a set of dark, cool loops overlie the lighter, lower-lying loops co-aligned with the bright \ion{Si}{4} loops. Case \#2 occurs below this set of highly variable loops, just to the north of the dark, presumably cold fibrils. IRIS spectra of the \ion{Si}{4} lines show extremely broad profiles, very strong blueshifts and intensity variations of a highly intermittent nature. Figure~\ref{fig:reconnect_si4} shows images and \ion{Si}{4} spectra at 08:56:29~UT at two locations (marked by blue and red crosses) lying in the flux emergence area. Both the images and the spectra suggest that this is the site of very strong reconnection-driven up and down flows. Reconnection is assumed here because in one location, marked by the blue cross at the southern edge of the dark bubble, we find upflow velocities of more that $200$~km$\,$s$^{-1}$ as well as enhanced red shifted emission of almost the same magnitude. A little bit further south, marked by the red cross in Figure~\ref{fig:reconnect_si4} (and covered by the dark overlying H$\alpha$ core fibrils in Figure~\ref{case3_mosaic}), we also find absorption dents in the \ion{Si}{4} 139.3~nm spectra probably caused by the existence of \ion{Ni}{2} 139.33 nm, showing that significant amounts of cool ($\sim 4000$~K) material exist above the energized transition region temperature loops \citep{2014Sci...346C.315P}. We note that the orientation of the newly emerging bubble described in case \#2 (given by the direction of separation of the opposite polarity magnetic feet in Figure~\ref{case3_mosaic}) is at an angle of some 90$^\circ$ to the apparently lower set of loops seen in the H$\alpha$ core and \ion{Si}{4}. Again, we interpret these highly dynamic and episodic loops as being energized by reconnection. As newly emerging flux rises from lower layers, it interacts and reconnects with the already existing loop system, heating and accelerating plasma to temperatures of order $100$~kK and supersonic velocities well above $100$~km$\,$s$^{-1}$. Above the interacting loops we find yet another set of loops containing cool absorbing plasma causing \ion{Ni}{2} absorption (seen diagonally crossing the H$\alpha$ core image in Figure~\ref{case3_mosaic}). However, it is difficult to establish a one-to-one correlation between a given emergence event (i.e. a dark bubble appearing in the chromospheric emission) and increased transition region activation: in addition to case \#2, we find at least two other similar emergence events in the immediate vicinity, i.e. within a radius of 5$\arcsec$, and within a timespan of 10 minutes of case \#2. Presumably there are many more occurring both before and after the events described here. 

\section{Discussion and conclusions}
\label{disc}

We have analysed small scale flux emergence in active region NOAA~11850 on September 25, 2013, using co-spatial and co-temporal SST, IRIS, and SDO observations. At the time of observations this was a region with vigorous flux emergence showing several energetic events such as Ellerman bombs or similar transition region explosions \citep{2013ApJ...774...32V,2014Sci...346C.315P}, rapidly varying transition region loop emission as well as the effects of granular scale emergence as reported here. In particular, we have observed lines formed throughout the solar atmosphere from the photosphere, up through the middle and upper chromosphere, the lower transition region and the corona to form an impression of how magnetic flux breaks through the photosphere and expands into the outer solar atmosphere. 

We confirm earlier findings reported in Papers~{\sc i} and {\sc ii} that a small scale (granular sized) cold magnetic bubble rises from the photosphere to the mid chromosphere in about 3.5 minutes, indicating a rise speed of 5~km\,s$^{-1}$. Morphologic and dynamic properties in the lower layers mimic those described in Paper~{\sc i}. This time, however, we have been able to reach further up as we detect a reaction to the passage of the bubble in upper chromosphere and transition region lines in both cases considered here. The upflow velocity of the magnetic bubble in the chromosphere and transition region is of order 5-10~km\,s$^{-1}$. We have inferred delays of 13 (case \#1) and 9 (case \#2) minutes between the first signs of flux emergence in the photosphere and the reaction of the \ion{Si}{4} lines. Such delays and velocities imply a transition region height of between 2700~km and 3900~km. While these heights are higher than the 1700--2200~km found in FAL models \citep[e.g.][]{1993ApJ...406..319F}, recent ``realistic'' 3D simulations that include flux emergence \citep[e.g.][]{2010ApJ...718.1070H,2008ApJ...679..871M,2009ApJ...702..129M} suggest that these numbers are reasonable. In fact, one expects the chromosphere to expand and push aside the corona, as flux enters the outer solar atmosphere \citep{2014ApJ...788L...2A}. This expansion will be either helped or hindered to some extent by the magnetic field pre-existing in the corona, as has already been studied by \citet[][and references cited therein]{1977ApJ...216..123H,2004A&A...426.1047A,2007ApJ...666..516G}. 

Strong intensity enhancements and energy release are observed whenever the new flux interacts with the ambient pre-existing loops.

\subsection{Possible mechanisms at work}

Emerging flux regions often have a complex field topology: as magnetic flux emerges and rises, pushing aside the previously existing hot coronal and transition region gas, we find that layers of cool plasma organized in fibrils accumulate as seen in Figures ~\ref{caso1_mosaic} and ~\ref{case3_mosaic}. This gas takes considerable time to drain sufficiently to become transparent, both in H$\alpha$ and at wavelengths $<25$~nm due to neutral hydrogen and helium absorption. New flux emergence events below cause new bubbles to rise into the upper atmosphere, where they interact with previously existing field. When the angle between these flux systems is large enough, heating of sufficient strength to raise temperatures to $10^5$~K or higher will ensue, aided by the lower densities of plasma found at greater heights. However, whether or not this heating is sufficient to propel the gas temperature to greater than $10^6$~K, creating low lying coronal plasma, is difficult to ascertain due to the high opacity of the cool fibrils (as evidenced by \ion{Ni}{2} absorption) overlying the reconnecting newly emerging flux.

Using opacities tabulated by \citet{2005ApJ...622..714A} and ionization states for hydrogen and helium from
\citet{1998A&AS..133..403M} we can calculate the optical path length through a typical cool fibril given estimates of the
temperature and density. We find that when assuming a temperature of $10^4$~K and a particle density of $10^{11}$~cm$^{-3}$ that
emission is reduced a factor 100 in $3\,200$~km in the AIA $17.1$~nm band, in $2\,300$~km in the $19.3$~nm band, and in only $800$~km in the $30.4$~nm band. Thus, typical H$\alpha$ core fibrils, such as those seen in Figure~\ref{case3_mosaic}, need only be of thickness a few thousand kilometers or less in order to attenuate the AIA signal in all coronal channels quite significantly.

Further evidence that significant amounts of cool material exist above both the region of flux emergence and the primary site of reconnection with overlying loops comes from \ion{Si}{4} spectra gathered in the vicinity of the emerging case \#2 bubble. Those spectra show absorption by the 
\ion{Ni}{2} 139.3~nm line which lies almost 100~km$\,$s$^{-1}$ to the blue of the \ion{Si}{4} line center. This absorption is presumably caused by the very dark, high fibrils seen in the H$\alpha$ core image just south of case \#2. 

Our interpretation of the observations is then that the active region chromosphere, transition region and corona is built up by the successive emergence of small granular scale bipoles proceeding much in the manner as sketched by \citet{2004ApJ...614.1099P}: 
The serpentine field, with undulations separated by some few Mm, rids itself of mass through reconnection, but still expands into the upper chromosphere containing sufficient mass to provide significant opacity in H$\alpha$ and in the hydrogen and helium continua below 20~nm. The chromospheric or coronal length of these field lines become comparable to the size of the active region as this process proceeds. Their orientation, which can change in time, reflects the (time-dependent) orientation of the large scale flux sheet impinging on the photosphere from below. The new magnetic field involved in the flux emergence process pushes the outer atmosphere upward, pushing cool material at much greater heights than in the quiet sun or in non-emerging active regions. Significant activity ensues as newly emerged flux interacts -- and eventually reconnects -- with previously emerged flux. This may be why slit-jaw movies of such an emerging flux region show the atmosphere to be riddled with short-lived brightenings that continue to occur over the course of the many hours and days it takes for the flux emergence to take place. But the coronal part of the new loops may not be visible since the entire flux emerging region is covered by several sets of overlapping loops, oriented at different angles, containing cool plasma that absorbs light at wavelengths $<30$~nm through bound free H~{\sc i}, He~{\sc  i}, and He~{\sc ii} transitions. 

\acknowledgments
Financial support by the Research Council of Norway through grants
208027/F50 and 'Solar Atmospheric Modelling', by the European Research
Council under the European Union's Seventh Framework Programme
\mbox{(FP7/2007-2013)} / ERC Grant agreement no.\ 291058, and by the
Programme for Supercomputing of the Research Council of Norway through 
grants of computing time are gratefully acknowledged. Financial support by the Spanish Ministerio de Econom\ia a
y Competitividad through grant ESP2013-47349-C6-1-R is gratefully
acknowledged. B.D.P. was supported by NASA under contract NNG09FA40C (IRIS), NNX11AN98G and NNM12AB40P. IRIS is a NASA small explorer mission developed and operated by LMSAL with mission operations executed at NASA Ames Research center and major contributions to downlink communications funded by ESA and the Norwegian Space Centre. This work benefited from discussions at the ISSI meetings on ``Heating of the magnetized chromosphere''. The Swedish
1~m Solar Telescope is operated by the Institute for Solar Physics of
Stockholm University in the Spanish Observatorio del Roque de los
Muchachos of the Instituto de Astrof\'{\i}sica de Canarias. This research has made use of NASA's Astrophysical Data System.

\bibliographystyle{aa}
\bibliography{jaime,solarrefs}

\clearpage

\end{document}